# QC-Automator: Deep Learning-based Automated Quality Control for Diffusion MR Images

Zahra Riahi Samani[1*], Jacob Antony Alappatt[1], Drew Parker[1], Abdol Aziz Ould Ismail[1], Ragini Verma[1]

[1]Diffusion and Connectomics in Precision Healthcare Research Lab (DiCIPHR), Department of Radiology, University of Pennsylvania, Philadelphia, PA+

**\* Correspondence:**
Zahra Riahi Samani
Zahra.RiahiSamani@PennMedicine.upenn.edu

**Abstract**

Quality assessment of diffusion MRI (dMRI) data is essential prior to any analysis, so that appropriate pre-processing can be used to improve data quality and ensure that the presence of MRI artifacts do not affect the results of subsequent image analysis. Manual quality assessment of the data is subjective, possibly error-prone, and infeasible, especially considering the growing number of consortium-like studies, underlining the need for automation of the process. In this paper, we have developed a deep-learning-based automated quality control (QC) tool, *QC-Automator*, for dMRI data, that can handle a variety of artifacts such as motion, multiband interleaving, ghosting, susceptibility, herringbone and chemical shifts. QC-Automator uses convolutional neural networks along with transfer learning to train the automated artifact detection on a labeled dataset of ~332000 slices of dMRI data, from 155 unique subjects and 5 scanners with different dMRI acquisitions, achieving a 98% accuracy in detecting artifacts. The method is fast and paves the way for efficient and effective artifact detection in large datasets. It is also demonstrated to be replicable on other datasets with different acquisition parameters.

## 1. Introduction

Diffusion MRI (dMRI) (Basser and Jones 2002, Assaf and Pasternak 2008) is now widely used to probe the microstructural properties of biological tissues, as well as the structural connectivity of the brain. dMRI is prone to different kinds of artifacts including motion, multiband interleaving, ghosting, susceptibility, herringbone and chemical shift (Wood and Henkelman 1985, Smith, Lange et al. 1991, Simmons, Tofts et al. 1994, Schenck 1996, Heiland 2008, Moratal, Vallés-Luch et al. 2008, Krupa and Bekiesińska-Figatowska 2015). If these artifacts remain undetected or insufficiently corrected, it could bias the results of subsequent analyses, weakening their interpretability (Bammer, Markl et al. 2003, Van Dijk, Sabuncu et al. 2012, Reuter, Tisdall et al.

2015). Thus, quality control (QC) is an essential step before dMRI goes into further processing like motion correction and tensor estimation (Bastiani, Cottaar et al. 2019).

QC is undertaken mostly by visual inspection prior to any processing or analysis, in order to assess the quality of the data. Based on this QC, appropriate corrections can be applied, or a decision can be made to exclude affected slices or volumes. This is very time consuming and challenging to undertake manually, especially in large datasets of dMRI data, a modality with inherently low signal to noise ratio. Furthermore, manual visual QC is subjective based on the level of sensitivity, expertise or even tolerance to fatigue of the QC expert, leading to high inter-rater variability (Victoroff, Mack et al. 1994). This warrants the need for automated QC methods, to limit the work of the QC expert to the inspection of slices that have been flagged by an automated algorithm. In this paper, we propose to design such an automated QC method to detect a wide range of artifacts that may occur individually or in combination, flagging affected slices for subsequent inspection. This can be applied prior to processing, as well as at any stage when the results of an analysis step need to be tested. This will help the user determine the presence of an artifact, and whether corrective steps need to be employed or the slices need to be excluded.

Some form of QC is present in the different artifact correction tools such as FSL EDDY (Andersson, Graham et al. 2016, Bastiani, Cottaar et al. 2019), DTI studio (Jiang, Van Zijl et al. 2006), DTIPrep (Oguz, Farzinfar et al. 2014) and TORTOISE (Pierpaoli, Walker et al. 2010). Such tools are usually limited to detecting and correcting the specific artifact that they have been designed for, mostly motion, and eddy current induced distortions (Liu, Zhu et al. 2015) (Kelly, Pietsch et al. 2016, Iglesias, Lerma-Usabiaga et al. 2017, Alfaro-Almagro, Jenkinson et al. 2018, Graham, Drobnjak et al. 2018). The results of these correction packages also need subsequent inspection to detect the presence of any remaining artifacts, making QC essential before and after these correction methods. However, these methods do not detect or correct for other prominent artifacts like ghosting, herringbone and chemical shifts, further underlining the need for a comprehensive QC paradigm, outside of these artifact correction packages.

While traditional feature-based machine learning methods can be considered as a natural choice for training artifact detection, these require careful feature selection, which can present a challenge considering the variety of artifacts, and noise, in dMRI data. This is further compounded by the fact that the same artifact may present differently across scanners/sites, making the feature-based learners site and scanner specific. Human QC experts rely on the brain's ability to identify and integrate patterns specific to artifacts in dMRI data to detect them. Deep learning tools, especially convolutional neural networks (CNN), that emulate human visual feature extraction in an automated manner, can be a very powerful tool for training an automated QC detector. The superior performance of CNN in many computer vision tasks, and in medical imaging, motivated us to use it to train an automated QC method for dMRI data.

In order to train a CNN that emulates human behavior, a large set of parameters need to be optimized during the training process, which in turn necessitates a high volume of training data (slices with artifacts, and slices of good brain tissue), and increased computational cost. Providing this huge volume of data is a challenging task, especially in medical imaging. In order to fulfill the requirement of a large amount of labeled data for training a deep CNN, transfer learning (Mazurowski, Buda et al. 2018) is used. Transfer learning involves taking a pre-trained CNN and re-training a subset of its parameters using a smaller amount of data to perform well on a new task (Mazurowski, Buda et al. 2018). As a result of this vast reduction in the number of parameters, transfer learning has the advantage of requiring less training time and computational cost. Pre-trained models have been applied successfully to various computer vision and medical imaging tasks, such as breast cancer diagnosis in digital breast tomosynthesis from mammography data (Samala, Chan et al. 2018, Samala, Chan et al. 2018), classification of radiographs to identify hip osteoarthritis (Xue, Zhang et al. 2017) or diagnosis of retinal diseases in retinal tomography images (Rampasek and Goldenberg 2018). As our sample size was limited due to the difficulty of manual QC labeling of dMRI data, we adopted a transfer learning approach in this paper.

A significant problem in artifact detection is that the same artifact may present differently across sites and scanners. In order to make the CNN insensitive to scanner and site differences, we use manually labeled datasets from different sites and scanners. In addition to this, we apply data augmentation techniques, that led to demonstrably improved results of CNN classifiers (Wang and Perez 2017). In this manner, classical image transformations, including rotating, cropping, zooming, and shearing, are applied on the original images to increase the heterogeneity of the sample, by providing a simulated variation of the original data. In the process, both heterogeneity and size of the sample are increased.

In summary, we present a CNN-based automated QC paradigm, called QC-Automator, to detect various artifacts in dMRI data, including motion, multiband interleaving, ghosting, susceptibility, herringbone and chemical shift. We will use transfer learning and data augmentation. The method will be trained and cross-validated on a large sample of expert-labeled images that combine dMRI data from multiple scanners.

## 2. Materials and Methods

Proposed method contains two CNN based classifiers, one for artifacts that manifest clearly in axial slices (e.g. ghosting), and one for artifacts that manifest in sagittal slices (e.g. motion). An input dMRI volume is converted into axial and sagittal slices and the slices are sent to the axial or sagittal classifier correspondingly. Finally, the slices in which artifacts are detected, by either of the two classifiers described above, are flagged and a slice-wise report is created.

We first describe the datasets that are used for training and testing, and describe the different artifacts in section 2.1. QC-Automator, is described next. The performance of a number of different CNN architectures are compared for their suitability to the problem of artifact detection and compared to traditional machine learning approaches using texture features. Additionally, we report the performance of the detectors on data of different acquisition protocols, that are not a part of the training set.

### 2.1. Database for training and testing QC-Automator

Our database for all the following experiments included data from 155 unique subjects across 5 different scanners and dMRI acquisition schemes. The details are reported in Table 1. The ground truth labels in this paper were provided by manual visual inspection. In order to reduce the manual labeling errors, QC was done by two experts with 2-8 years of experience. The labels were binarized, in order to create a classifier which categorizes images as 'artifact free' or 'artifactual'.

The artifacts labeled from these datasets were divided into six categories, motion, multiband interleaving, ghosting, susceptibility, herringbone and chemical shifts. These six categories of artifacts manifested differently in the images; thus, the QC experts inspected axial slices for herringbone, chemical shift, susceptibility, and ghosting artifacts, while they inspected sagittal slices for motion and multiband interleaving artifacts.

To exclude slices that capture the periphery of the brain which contain mostly background voxels, we excluded sagittal slices which were entirely outside of the brain and 5 sagittal slices starting from the left and right edges of the brain. We excluded 5 axial slices inferior to the superior surface of the skull, as well as slices superior to the skull and inferior to the cerebellum, as they represent non-brain tissue. Overall, ~132000 axial slices and ~200000 sagittal slices were annotated as either artifactual or artifact-free. Details are reported in Table 2. Figure 1 shows representative examples of the artifacts that were annotated, based on the view used.

*Table 1- The acquisition parameters across our datasets*

| Datasets | # subjects | b-values (s/mm$^2$) | #repeated acquisitions | #b=0 Images | #Weighted Gradients | TR (ms) | TE (ms) |
|---|---|---|---|---|---|---|---|
| Dataset-1 | 30 | 1000 | 2 | 1 | 32 | 8000 | 51 |
| Dataset-2 | 32 | 1000 | 2 | 7 | 30 | 6500 | 84 |
| Dataset-3 | 17 | 300, 800, 2000 | 1 | 9 | 108 | 4300 | 75 |
| Dataset-4 | 31 | 1000 | 1 | 7 | 64 | 8100 | 82 |
| Dataset-5 | 57 | 1000 | 1 | 1 | 30 | 11000 | 76.4 |

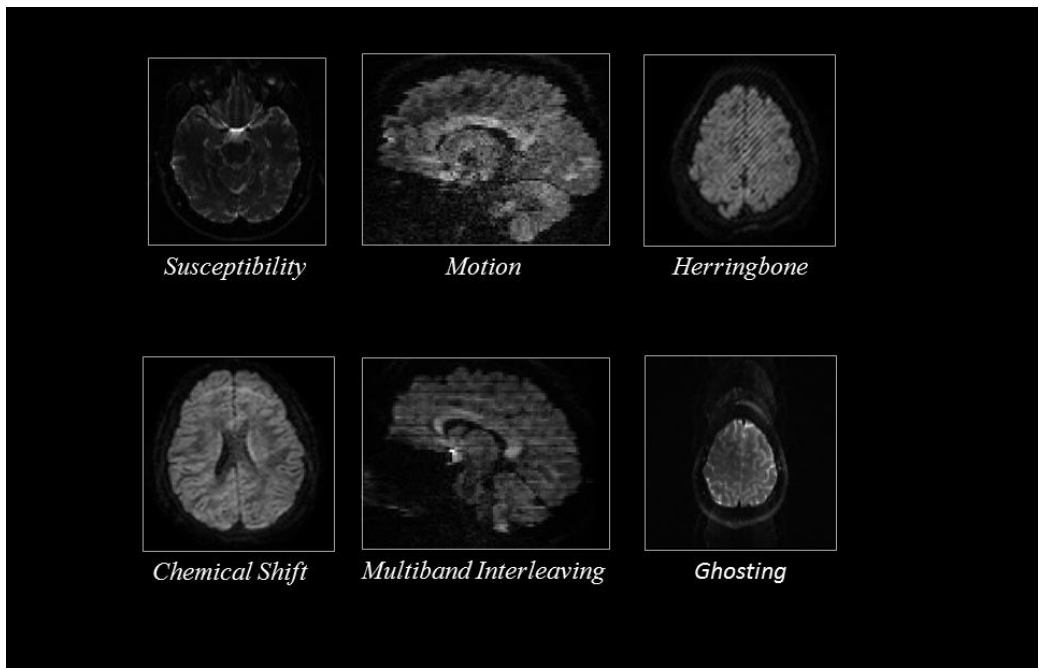

*Figure 1- Representative slices of the different artifacts that the QC-Automator was trained to detect.*

*Table 2- Distribution of different types of artifacts in our dataset*

| Artifact Type | Slice View | Total Samples |
|---|---|---|
| Herringbone | Axial | 120 |
| Chemical Shift | Axial | 1054 |
| Susceptibility | Axial | 442 |
| Ghosting | Axial | 11619 |
| Motion | Sagittal | 21436 |
| Multiband Interleaving | Sagittal | 4017 |
| Total-Artifact | Axial | 13235 |
| Total-Artifact | Sagittal | 25453 |
| Total-Artifact-Free | Axial | 118641 |
| Total- Artifact-Free | Sagittal | 179911 |

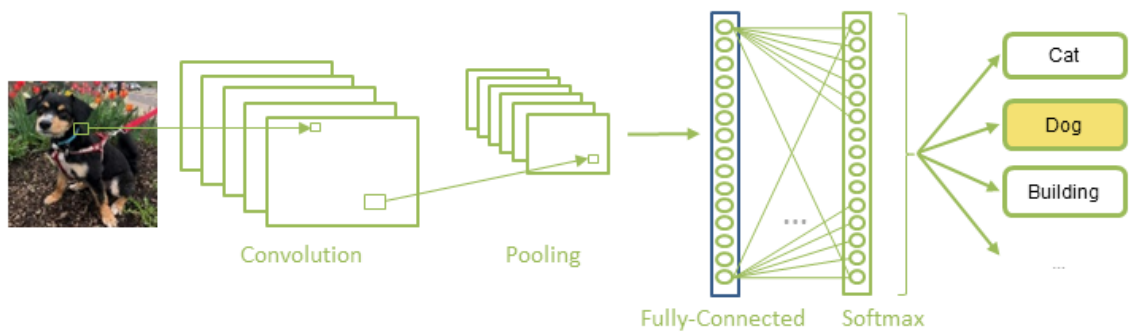

*Figure 2- A typical architecture of a CNN: A set of convolution and pooling layers with successive fully connected and softmax layer.*

## 2.2. Convolutional Neural Networks
### 2.2.1. An overview

CNNs are a special kind of artificial neural network that are composed of a set of convolutional and pooling layers in their architectures (Figure 2). Convolutional layers are designed to detect certain local features throughout the input image; they perform a convolution operation to the input image and pass the result to the next layer, a pooling layer, which reduces the dimensionality of the data by combining the outputs of a set of neurons into a single one, via a max or average operation. A sequence of convolutional and pooling layers is followed by some successive fully connected layers, in which all the neurons in a prior layer are connected to all the neurons in the next layer. Finally, a softmax, or regression layer tags the data with the desired output label. (Krizhevsky, Sutskever et al. 2012).

Various CNN architectures have been proposed in literature. The VGG (Simonyan and Zisserman 2014) networks, along with the earlier AlexNet (Krizhevsky, Sutskever et al. 2012), are the most basic architectures which follow the traditional layout of CNNs as shown in Figure 2. ResNet (He, Zhang et al. 2016), Inception (Szegedy, Liu et al. 2015) and Xception (Chollet 2017) are newer architectures. While ResNet introduces residual networks that make some connections between non-consecutive layers in very deep networks, Inception uses a module that performs different transformations over the same input in parallel and concatenates their results. Xception, on the other hand, is based on separating cross-channel and spatial correlations. Each of these architectures convey their own unique advantages and pitfalls, warranting a comparison of the performance of different CNN architectures.

### 2.2.2. Transfer Learning

To train CNNs, a large number of parameters need to be optimized, which in turn requires a large amount of computational power and labeled training data (in our case, the database of artifactual and artifact-free slices). Manually labeling data, however, is a time-consuming process, and with the limited number of datasets in medical imaging, it may not be possible to create a large and heterogeneous enough database to train a CNN from scratch for a given task. To overcome these issues, transfer learning methods have been proposed in which, an existing CNN network/architecture, pre-trained on a certain task, is adapted to a new task and CNN parameters are adjusted for a few layers of the network.

In general, the early layers of a CNN learn low-level features, which are applicable to most computer vision tasks, while the subsequent layers learn high-level features that are mostly application-specific. Therefore, adjusting the last few layers of an existing CNN architecture is usually sufficient for transfer learning (Tajbakhsh, Shin et al. 2016).

The efficiency of the transfer a learning method depends on the similarity between the images of database the selected CNN architecture was trained on and the images of the database that we want to transfer the CNN to. Although the heterogeneity between the images used in the pre-trained CNNs (see section 2.2.1) and medical imaging databases is considerable, an extensive study on medical imaging data has demonstrated that adjusting the parameters of an existing, pre-trained CNN, is as effective as training a CNN from scratch while being more robust to the size of training data (Tajbakhsh et al. 2016) and requiring significantly less computational power.

As providing the manual labels for dMRI data is a time consuming and laborious task, our sample size was limited and insufficient to train a CNN from scratch. In this paper we used transfer learning, to create QC-Automator, described in detail in the following section.

### 2.3. Creation of QC-Automator

Figure 3 shows the pipeline of the proposed approach. The artifacts detected by QC-Automator are motion, multiband interleaving, ghosting, susceptibility, herringbone and chemical shifts. As different artifacts manifested more clearly either in the axial or sagittal view, QC-Automator consisted of two detectors: the "axial detector" which detected artifacts that presented better axially (herringbone, chemical shift, susceptibility and ghosting) and the "sagittal detector" which detected artifacts that presented in the sagittal plane (motion and multiband interleaving artifacts).

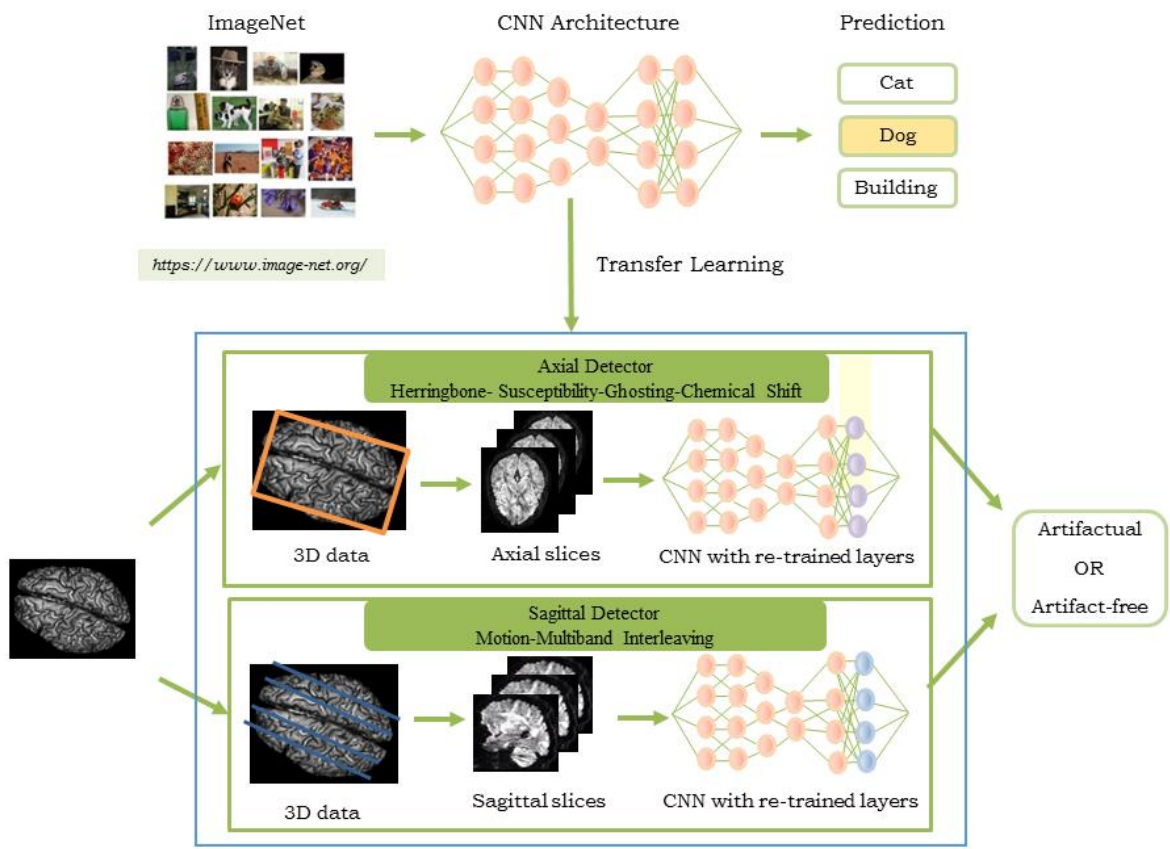

*Figure 3- Pipeline of the proposed approach for the QC-Automator: (Top) CNN pre-trained on ImageNet to obtain parameters used for transfer learning, where the last layer of the network was re-trained with our dataset of manually labeled artifactual and artifact-free data. The process was replicated to create the axial (middle row) and the sagittal detector (bottom row). The blue box represents the QC-Automator. Given an input image (left), both the axial and sagittal detectors are applied to it and the status of each slice as artifact-free or artifactual is predicted.*

For creating training samples, every dMRI volume was converted to axial and sagittal slices, and was assigned manual labels (see Section 2.1 for details). The two detectors were fed a new dMRI volume, in order to determine the slices that manifest with artifacts.

### 2.3.1. Transfer Learning and Data Augmentation for QC-Automator

In the proposed architecture, the detectors were CNN based and applied transfer learning to adapt existing knowledge, obtained from a large database of labeled training samples in other domains, to our problem of artifact detection ( see Section 2.2.2 for details). The transfer learning process consists of two main steps: selection of the pre-trained model, and applying the pre-trained model to the new domain. To select an optimal pre-trained model for our axial and sagittal detectors, we compared the performance of four different pre-trained CNN architectures namely, VGGNet, ResNet, Inception, and Xception. These CNN architectures were pre-trained on ImageNet (Russakovsky, Deng et al. 2015), which is the most popular public dataset with a very large amount of labelled images across various number of classes. This makes these architectures capable of learning generic features from images, making them good feature extractors for a variety of classification tasks.

To implement transfer learning, we removed the top layer of a pre-trained CNN and replaced it with a fully connected layer with 256 neurons, followed by a softmax layer which performs the classification between two classes (artifact-present vs artifact-free). All parameters of CNN architectures were fixed except those in the newly added layer, which were re-trained with the augmented manually-labeled artifactual and non-artifactual data described above.

Data augmentation techniques are strategies that enable a significant increase in the diversity and size of data available for training, without collecting new data. They perform, different image transformations to provide a simulated variation of the original data for training. We performed extensive data augmentation of the manually labeled data by applying horizontal and vertical translations, rotations, zooming, shearing and flipping of the original slices. This was undertaken to increase the sample size of the labeled dataset, as well as to increase the heterogeneity of the data.

### 2.3.2. Training QC-Automator

The two classifiers were trained using the first 3 datasets, by passing axial and sagittal slices of the brain along with ground-truth labels (artifactual, or artifact-free) For both classifiers, the intensity values for each slice were normalized to have zero center and unit variance as calculated by the value subtracted by the mean and divided by the standard deviation. Training was done for 20 epochs using the RMSprop optimizer with a learning rate of $2\times10^{-4}$ and a cross entropy loss function. The network structure was implemented in Python, using Keras with Tensorflow as the backend (Python 2.7, Keras 2.0.8, Tensorflow 1.3.0).

### 2.3.3. Slice-based and Volume-based reports

QC-Automator was designed to produce a report of the presence or lack of artifacts in individual slices in a diffusion-weighted image. However, an alternate way of reporting such information is based on the presence or absence of artifacts in an entire volume. To this end, we used a slice-count threshold to label a volume as "artifactual". If QC-Automator found that a given volume contained more artifactual slices than the slice-count threshold, it would flag this volume as artifactual. While choosing low values of slice-count threshold could lead to over detection, choosing high values for threshold could lead to higher chances of missing artifacts by not flagging a volume. We chose different slice-count values for the threshold from 1 to 10 in order to find an optimal threshold.

To summarize the pipeline of the QC-Automator, an input dMRI volume was sliced axially and sagittally, and the respective slices were sent to the axial classifier, or the sagittal classifier. The presence of an artifact was detected by either of the two classifiers, and the artifactual slices were flagged in a slice-wise report.

### 2.4. Evaluating the performance of QC-Automator

The following measures were used to evaluate the performance of QC-Automator:

$$Precision = \frac{TP}{TP + FP}$$

$$Recall = \frac{TP}{TP + FN}$$

$$Accuracy = \frac{TP + TN}{TP + TN + FP + FN}$$

where true positive (TP) represents the number of cases correctly recognized as artifactual, false positive (FP) represents the number of cases incorrectly recognized as artifactual, true negative (TN) represents the number of cases correctly recognized as artifact-free and false negative (FN) represents the number of cases incorrectly recognized as artifact-free.

#### 2.4.1. Comparing CNN Architectures to Traditional Methods

The performance of the four different architectures (VGGNet, ResNet, Inception and Xception) were evaluated on the first three datasets in a 5-fold cross-validation setting. The slices from the first three datasets were shuffled and randomly partitioned into 5 equally sized subsamples. For each run, a single subsample was retained as test data, while the remaining 4 subsamples were used as training data, and this process was repeated for all 5 subsamples. This cross-validation process was run on the "axial" and "sagittal" detectors separately.

We compared our approach with three traditional feature extraction and learning methods: Gabor features, Zernike moments and Local Binary Patterns. Gabor features are constructed from the responses of applying Gabor filters made on several frequencies (scales) and orientations (Manjunath and Ma 1996). We applied Gabor filters with 4 directions and 4 scales. Zernike moments are a global image feature constructed by projecting the image onto Zernike Polynomials, which are a set of orthogonal basis functions mapped to the unit circle in different orders and repetitions (Khotanzad and Hong 1990, Revaud, Lavoué et al. 2009). We applied Zernike moments with order 4 and repetition 2. Local Binary Patterns are a non-parametric method to detect local structures of images by comparing each pixel with its neighboring pixels (Ojala, Pietikainen et al.

2002, Huang, Shan et al. 2011). We applied the aforementioned features in combination with random forest classifiers using the same cross-validation scheme described above.

In order to investigate the CNN classifiers and the traditional filters individually, we designed two more experiments. In the first experiment, the outputs of Gabor filters on images were fed into a fully connected layer with 256 neurons followed by a dropout and a softmax layer. In the second experiment, we applied principal component analysis (PCA) on the output of final convolutional layer of the CNN. We kept enough principal components to cover 98% of the variation in the data, and fed them into an SVM classifier.

### 2.4.2. Evaluating Performance on New Datasets

We performed the following experiments to evaluate whether the artifact detection is replicable to other dMRI datasets with different acquisition protocols. We tested the applicability of the QC-Automator on two new datasets, Dataset 4 and Dataset 5 (details are in Table 1). These datasets contained a variety of artifacts, encompassing all those that the detector was trained to detect, but were acquired with different scanning parameters. The training was done using the first three datasets, while the performance was evaluated using data from the fourth and fifth datasets.

In addition, we investigated whether generalizability in performance across datasets was improved by retraining QC-Automator after adding a small subsample (10%) of the new datasets to the training set, to see if incorporating samples could improve the accuracy, precision and recall of the classifier versus application to a hitherto unseen dataset. We performed two different experiments by adding data from Dataset 4 and Dataset 5 to the original training set, separately.

## 3. Results

### 3.1. Comparison across CNN Architectures and Traditional Methods

Table 3 and Table 4 show the performance of our artifact detection method, using different architectures. As VGGNet outperformed other architectures, it was selected as the architecture of choice for QC-Automator. Using VGGNet, we obtained 98% accuracy for all artifacts in both the axial and sagittal detectors. Precision and recall values are reported accordingly. Representative instances of the true and false detections for QC-Automator are shown in Figure 4 and Figure 5.

*Table 3- The result of different CNN architectures in detecting artifact type 1.(Axial Detector).*

|  | Accuracy | Precision | Recall |
|---|---|---|---|
| **VGG 16** | 0.98 | 0.97 | 0.91 |
| **Resnet 50** | 0.89 | 0.82 | 065 |
| **Inception V3** | 0.96 | 0.89 | 0.82 |
| **Xception** | 0.96 | 0.88 | 0.82 |

*Table 4- The result of different CNN architectures in detecting artifact type 2.(Sagittal Detector)*

|  | Accuracy | Precision | Recall |
|---|---|---|---|
| **VGG 16** | 0.98 | 0.92 | 0.91 |
| **Resnet 50** | 0.98 | 0.91 | 0.78 |
| **Inception V3** | 0.98 | 0.90 | 0.67 |
| **Xception** | 0.99 | 0.92 | 0.82 |

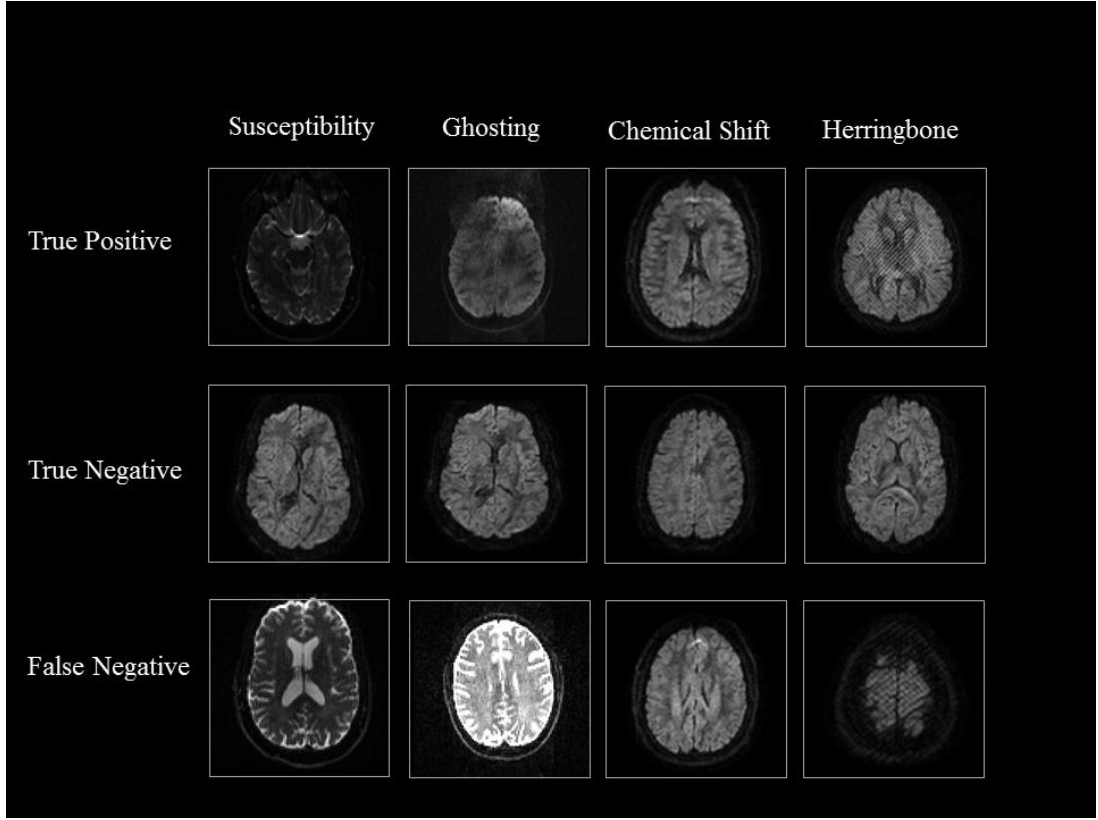

*Figure 4- Results of Axial Detector: Representative slices of correctly and incorrectly classified slices are presented.*

Table 5 and Table 6 compare our method with traditional pattern recognition approaches including Gabor filters, Zernike moments and local binary patterns in combination with random forest classifiers. As seen, VGGNet outperformed the traditional methods. Table 7 shows the result of applying Gabor filters to a fully connected layer and Table 8 shows the results of performing SVM on top of VGGNet final convolutional layer features after PCA. Although Gabor filters and SVM classifiers could achieve high accuracy (87% and 91% for axial detector), the value of precision and recall was poor compared to our method using CNNs, showing that our transfer learning approach outperformed traditional SVM classifiers and Gabor filters for this task.

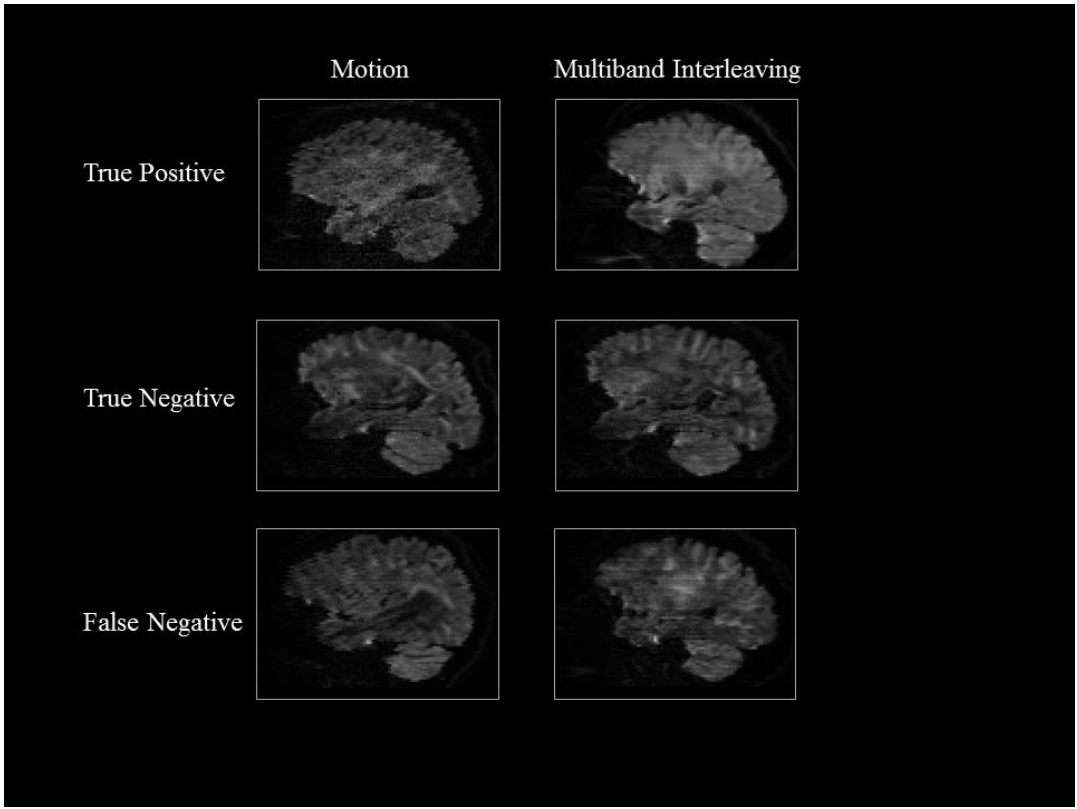

*Figure 5- Results of Sagittal Detector Representative slices of correctly and incorrectly classified artifactual slices.*

*Table 5- Results of different texture features in detecting artifact type 1.(Axial Detector).*

|  | **Accuracy** | **Precision** | **Recall** |
|---|---|---|---|
| **Gabor 32** | 0.91 | 0.89 | 0.87 |
| **Zernike Moments** | 0.87 | 0.58 | 0.19 |
| **Local Binary Patterns** | 0.83 | 0.85 | 0.12 |

*Table 6- Results of different texture features in detecting artifact type 2.(Sagittal Detector)*

|  | **Accuracy** | **Precision** | **Recall** |
|---|---|---|---|
| **Gabor -32** | 0.98 | 0.96 | 0.48 |
| **Zernike Moments** | 0.97 | 0.45 | 0.55 |
| **Local Binary Patterns** | 0.97 | 0.40 | 0.52 |

Volume-wise results for VGGNet are reported in Table 9 and Table 10. As seen, we obtained 96% accuracy for our axial detector at a slice-count threshold of 3 slices, and 98% accuracy for our sagittal detector at a slice-count threshold of 7 slices. Correspondingly, recall values were 98% and 95% for the volume-wise axial and sagittal detectors. This means we only missed 2% of volumes that contain artifacts manifesting in axial view and 5% of sagittal ones.

*Table 7-The Result of Gabor filter combined with fully connected layers*

| Gabor Filters- Fully Connected | Accuracy | Precision | recall |
|---|---|---|---|
| Axial Detector | 0.87 | 0.37 | 0.35 |
| Sagittal Detector | 0.90 | 0.30 | 0.46 |

*Table 8- The Result of feeding CNN features to Support vector machines*

| CNN-SVM | Accuracy | Precision | recall |
|---|---|---|---|
| Axial Detector | 0.91 | 0.94 | 0.85 |
| Sagittal Detector | 0.87 | 0.93 | 086 |

*Table 9-QC- Automator Volume-wise result- Axial Detector*

| Threshold | Accuracy | Precision | Recall |
|---|---|---|---|
| Threshold=1 | 0.92 | 0.86 | 0.99 |
| Threshold=3 | 0.96 | 0.94 | 0.98 |
| Threshold=5 | 0.94 | 0.97 | 0.90 |
| Threshold=7 | .0.87 | 0.97 | 0.74 |
| Threshold=10 | 0.84 | 0.69 | 0.67 |

*Table 10- QC- Automator Volume-Wise result- Sagittal Detector*

| Threshold | Accuracy | Precision | Recall |
|---|---|---|---|
| Threshold=1 | 0.74 | 0.64 | 0.97 |
| Threshold=3 | 0.90 | 0.79 | 0.96 |
| Threshold=5 | 0.97 | 0.87 | 0.95 |
| Threshold=7 | 0.98 | 0.94 | 0.95 |
| Threshold=10 | 0.98 | 0.97 | 0.95 |

### 3.2. Performance on new datasets

These experiments were undertaken in order to evaluate whether the artifact detection is replicable to other datasets acquired through different imaging protocols. The detectors were trained on the first 3 datasets and tested on the fourth and fifth datasets. Their performance is reported in Table 11 and Table 12. For Dataset 4, the accuracy of detecting artifacts through the axial detector decreased by 7% comparing to the previous results in Table 3. There was also a 14% decrease in the accuracy of the sagittal artifact detector, as compared to Table 4. For Dataset 5, the value of accuracy dropped by 7% and 11% for the axial and sagittal detectors, respectively.

In order to see if these results could be improved, we evaluated the results of adding a small percentage of the new datasets to the original training data, to acclimatize the deep learner to new scanning parameters. We added a small subset (10% of each whole dataset) from the fourth and fifth datasets to the original training set, the results of which are displayed in Table 13 and Table 14. It can be seen that we attained a higher accuracy, recall and precision than those of the previous experiment (Table 11 and Table 12). Results were in the range of 90% recall for both new datasets, demonstrating that we missed less than 10% of artifacts. we provided an example of a false positive case for this experiment in Figure 6.

*Table 11- Results of applying the QC-automator to the fourth dataset*

|  | Accuracy | Precision | Recall |
|---|---|---|---|
| **Artifact type-1-Axial** | 0.91 | 0.75 | 0.81 |
| **Artifact type 2-sagittal** | 0.84 | 0.70 | 0.79 |

*Table 12- Results of applying the QC-automator to the fifth dataset*

|  | Accuracy | Precision | Recall |
|---|---|---|---|
| **Artifact type-1-Axial** | 0.91 | 0.91 | 0.71 |
| **Artifact type 2-sagittal** | 0.87 | 0.75 | 0.69 |

*Table 13- Results of applying the QC-automator on the fourth dataset, after adding small subsample (10%) data from the fourth dataset to the training set*

|  | Accuracy | Precision | Recall |
|---|---|---|---|
| **Artifact type-1-axial** | 0.94 | 0.87 | 0.91 |
| **Artifact type 2-sagittal** | 0.95 | 0.84 | 0.90 |

*Table 14 Results of applying the QC-automator on the fifth dataset, after adding after adding small subsample (10%) from the fifth dataset to the training set*

|  | Accuracy | Precision | Recall |
|---|---|---|---|
| **Artifact type-1-axial** | 0.89 | 0.82 | 0.91 |
| **Artifact type 2-sagittal** | 0.94 | 0.84 | 0.94 |

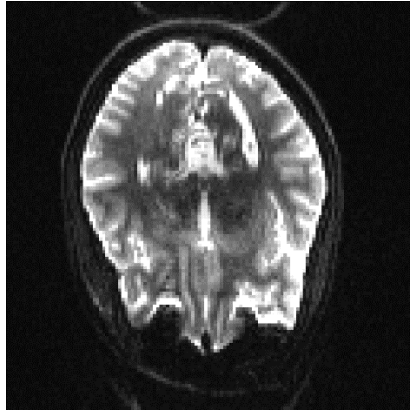

*Figure 6- A sample of False Positive slice for Dataset 4: The slice contains aliasing artifact. Our expert labelled it as artifact-free one. But our QC-Automator caught it as it contained a similar pattern to ghosting artifact.*

## 4. Discussion

In this paper, we created an automated QC method, QC-Automator, using CNN and transfer learning, via data augmentation on a manually labeled dataset encompassing several scanners and dMRI acquisition parameters. We demonstrated the ability of QC-Automator to distinguish between artifactual slices from artifact-free ones, as well as its performance across different acquisitions from multiple sites. Given a diffusion MRI volume, the QC-Automator was able to flag slices based on the presence of several artifacts, including motion, multiband interleaving, ghosting, susceptibility, herringbone and chemical shift. The flagged slices can be manually inspected to determine if the corresponding volume would be safe to use for further analysis for a given study.

The QC-Automator consisted of two classifiers: one for all artifacts that manifest in the axial view (namely herringbone, chemical shift, susceptibility and ghosting), and one for artifacts that manifest in the sagittal view (namely motion and multiband interleaving). For both the classifiers, VGGNet performed better than Inception, ResNet and Xception, based on the comparison of transfer learning results for various architectures (Table 3 and Table 4). This might be because of the uniform structure of VGGNet, which uses consecutive layers of 3x3 filters and max pooling, with each successive layer detecting features at a more abstract, semantic level than the layer before. Residual networks introduce connections between layers at different resolutions, which results in a jump in the semantic abstraction. Inception and Xception networks compute and concatenate multiple different transformations over the same input. These architectures are more complex and did not perform as well on our data.

The representative slices in Figure 4 and Figure 5 demonstrate that our method correctly classified artifacts in different slices of the brain. Despite having correctly classified most artifacts, the QC-Automator also incorrectly flagged some artifactual slices as artifact-free, and we inspected some of these examples. We hypothesize that the false negative case for ghosting (see Figure 4) happened because the pattern of ghosting was particularly faint in this specific slice. For herringbone, chemical shift and susceptibility artifacts, our classifier successfully labeled multiple slices of the given volume as artifactual, but sometimes failed to label slices where the artifact was less prominent (see Figure 4). Thus, although our classifier failed to correctly label some artifactual slices, it was able to capture adjacent slices where the artifact was more prominent. As the pattern of artifact is more visible in susceptibility, herringbone and chemical shift, we believe we can get better performance by adding more training data for other artifacts in the future.

The transfer-learning-based approach presented in this paper performed better than Gabor filters, Zernike moments and Local binary patterns in combination with random forest classifiers (Table 5 and Table 6). Gabor filters performed better than Zernike moments and Local binary patterns. The fact that Gabor filters analyze the presence of specific frequencies in specific directions of localized regions in the input image might explain this result. These specific frequencies can capture the edge patterns in motion and multi-band interleaving, the checkerboard pattern in herringbone and chemical shift, and the curves visible in the background of ghosting artifacts. While Gabor filters had the best performance of the three, the precision and recall were still poor compared to VGGNet. Zernike and local binary patterns on the other hand, look for patterns of intensity. This is enough for detecting high intensities but fail to find patterns of edges and curves. However, the performance of these methods is bound to the quality of the features, which need human experts to hand-craft them manually. The fact that the Gabor features did well lends support to the notion that most of our features were discriminated in the early layers of the CNN, and thus the transfer learning approach, which consists of adjusting only last layer, performed well. The proposed approach also performed better than Gabor filters in combination with fully connected layer neural networks (Table 7) and it performed better than Support Vector Machine classifiers (Table 8). This indicates that VGGNet is a good choice both as a feature extractor and as a classifier.

As an alternative to the slice-wise report, we also measured the performance of QC-Automator when reporting the presence of artifact in an entire volume (Table 9 and Table 10). This way of reporting is easier for the human analyst to interpret, than a flat list of bad slices. In this manner, a volume was labeled as artifactual if it has more artifactual slices than a certain threshold. For the axial detector, we observed 99% recall with a lower threshold, meaning we detected 99% of volumes containing artifacts. However, the precision was 86% at this point, implying that we over-detected in 14% of cases. As we increased the threshold, the precision improved and the recall decreased as the detector missed some artifactual volumes. Optimal results appeared at Threshold of 3 slices, with precision of 95% and recall of 98%. For our sagittal detector, however, the optimal threshold was higher. We got 97% recall at slice-count threshold of 1 slices, while precision was poor at this point (64%). As we increased the threshold, precision improved and recall dropped. The optimal point was at a slice-count threshold of 10 slices, as it had the highest values of recall and precision. This difference in the optimal thresholds between the detectors might be because of the nature of motion artifacts, which are generally visible in more than one sagittal slices in a volume. Overall, QC-Automator only missed 2% of artifactual volumes which contain artifact in axial view and 5% of volumes with artifact present in sagittal view.

Furthermore, the framework was tested on how well it performs on acquisitions from different scanners. Evaluation was performed by training on 3 datasets and testing on the two remaining datasets, which had different acquisition parameters compared to the 3 training datasets.

The corresponding results for Dataset 4 and Dataset 5 are shown in Table 11 and Table 12. It indicates that, we could achieve high accuracy (90% approximate average), however, the values of precision and recall decreased.

To inspect the decrease in precision and recall, we added small subsample of the fourth and fifth datasets to the training set which covers 10% of the dataset (see Section 3.2). In this experiment, we achieved higher accuracy, close to the intra-dataset experiment (Table 3 and Table 4). The value of precision and recall also increased substantially for both detectors in both datasets (approximately 10% for Dataset 4 and 20% for Dataset 5). This suggests that adding a small subsample of the new datasets to the original training set, could decrease the false detection. As seen, in this experiment we achieved a higher value of recall, around 90% for both datasets, showing that QC-Automator had low chance of missing an artifact, while staying in the range of 85% for precision. This indicates that there were some artifact-free slices that are detected as artifactual representing that our method over-detected in some cases. However, considering the nature and purpose of QC, a false positive is favorable to false negative, as we do not want to miss an artifact. To summarize, by adding a subsample of the new datasets to the original training set, a drastic increase in recall was observed, giving us reason to believe that the classifier could be gradually improved to reach the same level of precision and recall as that of the intra-dataset experiment. This means that with a little effort we can apply our classifier to a new dataset.

Finally, we inspected the false positives of the cross-dataset experiment which uncovered another potential cause of false positive; an error in labelling of the data. As it can be seen in Figure 6, there was an aliasing artifact inside the slice, despite the fact that our QC expert had labelled that slice as artifact-free. However, our classifier detected them as artifactual slices possibly due to the fact that this slice had similar patterns to the ghosting artifact. The fact that QC-Automator was able to detect such artifacts, despite potential mislabeling in the training dataset, indicates the high performance of the detectors.

Despite the impressive results of QC Automator, there is still room for improvement, such as by adding more training data. We trained our classifier on 3 datasets acquired on different scanners with varying fields of view and gradient sampling schemes and tested our classifier on two other datasets, again of different acquisition sequences. We observed a high accuracy in our cross-dataset experiment, however there was a decrease in the precision and recall implying higher rates of false detection. We believe that this issue can be solved by adding small subsamples of the target dataset so the training set so classifier can gradually get improved over-time with seeing more data.

The ground truth labels in this paper were provided manually. Artifacts manifest differently in different slices, from very subtle to clearly visible patterns. The subjectivity of manual visual inspection in our case was lowered by labeling using two experts, with varying degrees of expertise. The labels were binarized into two classes to create a classifier to categorize images as 'artifact-free' or 'artifactual'. If an objective 'artifact severity' threshold can be determined through characterization of artifacts, it might provide a better alternative to the use of binary labels.

Overall, the QC-Automator can gain from large training samples, limited by the effort and quality of manually labeling data on different artifacts. Given the recent progress in deep networks, and further advances in GPU hardware, the accuracy of convolutional neural nets is expected to further improve in the future. That provides the potential for better quality control tools.

## 5. Conclusion

In summary, QC-Automator is a deep learning-based method for quality control of diffusion MRI data that is able to detect a variety of artifacts. Quality control is a well-suited task for convolutional neural networks. The difficulty in obtaining huge amounts of expert-labeled dMRI data to train a CNN is alleviated by using transfer learning, and data augmentation. The proposed approach achieves superior performance with respect to pattern recognition methods and is considerably faster and less computationally expensive in comparison to purely learning-based approaches with neural networks. We demonstrated that our method achieves high accuracy and generalizes well to other datasets, different from those used for training. This artifact detector enhances analyses of dMRI data by flagging artifactual slices. This substantially reduces the effort and time of human experts and allows for an almost instantaneous access to clean dMRI data.

## References

Alfaro-Almagro, F., M. Jenkinson, N. K. Bangerter, J. L. Andersson, L. Griffanti, G. Douaud, S. N. Sotiropoulos, S. Jbabdi, M. Hernandez-Fernandez and E. Vallee (2018). "Image processing and Quality Control for the first 10,000 brain imaging datasets from UK Biobank." Neuroimage 166: 400-424.
Andersson, J. L., M. S. Graham, E. Zsoldos and S. N. Sotiropoulos (2016). "Incorporating outlier detection and replacement into a non-parametric framework for movement and distortion correction of diffusion MR images." NeuroImage 141: 556-572.
Assaf, Y. and O. Pasternak (2008). "Diffusion tensor imaging (DTI)-based white matter mapping in brain research: a review." Journal of molecular neuroscience 34(1): 51-61.
Bammer, R., M. Markl, A. Barnett, B. Acar, M. Alley, N. Pelc, G. Glover and M. Moseley (2003). "Analysis and generalized correction of the effect of spatial gradient field distortions in diffusion-weighted imaging." Magnetic Resonance in Medicine: An Official Journal of the International Society for Magnetic Resonance in Medicine 50(3): 560-569.
Basser, P. J. and D. K. Jones (2002). "Diffusion-tensor MRI: theory, experimental design and data analysis–a technical review." NMR in Biomedicine: An International Journal Devoted to the Development and Application of Magnetic Resonance In Vivo 15(7-8): 456-467.
Bastiani, M., M. Cottaar, S. P. Fitzgibbon, S. Suri, F. Alfaro-Almagro, S. N. Sotiropoulos, S. Jbabdi and J. L. Andersson (2019). "Automated quality control for within and between studies diffusion MRI data using a non-parametric framework for movement and distortion correction." NeuroImage 184: 801-812.
Chollet, F. (2017). "Xception: Deep learning with depthwise separable convolutions." arXiv preprint: 1610.02357.
Graham, M. S., I. Drobnjak and H. Zhang (2018). "A supervised learning approach for diffusion MRI quality control with minimal training data." NeuroImage.
He, K., X. Zhang, S. Ren and J. Sun (2016). Deep residual learning for image recognition. Proceedings of the IEEE conference on computer vision and pattern recognition.
Heiland, S. (2008). "From A as in Aliasing to Z as in Zipper: Artifacts in MRI." Clinical neuroradiology 18(1): 25-36.

Huang, D., C. Shan, M. Ardabilian, Y. Wang and L. Chen (2011). "Local binary patterns and its application to facial image analysis: a survey." IEEE Transactions on Systems, Man, and Cybernetics, Part C (Applications and Reviews) 41(6): 765-781.

Iglesias, J. E., G. Lerma-Usabiaga, L. C. Garcia-Peraza-Herrera, S. Martinez and P. M. Paz-Alonso (2017). Retrospective head motion estimation in structural brain MRI with 3D CNNs. International Conference on Medical Image Computing and Computer-Assisted Intervention, Springer.

Jiang, H., P. C. Van Zijl, J. Kim, G. D. Pearlson and S. Mori (2006). "DtiStudio: resource program for diffusion tensor computation and fiber bundle tracking." Computer methods and programs in biomedicine 81(2): 106-116.

Kelly, C., M. Pietsch, S. Counsell and J.-D. Tournier (2016). Transfer learning and convolutional neural net fusion for motion artefact detection. Proc. Intl. Soc. Mag. Reson. Med.

Khotanzad, A. and Y. H. Hong (1990). "Invariant image recognition by Zernike moments." IEEE Transactions on pattern analysis and machine intelligence 12(5): 489-497.

Krizhevsky, A., I. Sutskever and G. E. Hinton (2012). Imagenet classification with deep convolutional neural networks. Advances in neural information processing systems.

Krupa, K. and M. Bekiesińska-Figatowska (2015). "Artifacts in magnetic resonance imaging." Polish journal of radiology 80: 93.

Liu, B., T. Zhu and J. Zhong (2015). "Comparison of quality control software tools for diffusion tensor imaging." Magnetic resonance imaging 33(3): 276-285.

Manjunath, B. S. and W.-Y. Ma (1996). "Texture features for browsing and retrieval of image data." IEEE Transactions on pattern analysis and machine intelligence 18(8): 837-842.

Mazurowski, M. A., M. Buda, A. Saha and M. R. Bashir (2018). "Deep learning in radiology: an overview of the concepts and a survey of the state of the art." arXiv preprint arXiv:1802.08717.

Moratal, D., A. Vallés-Luch, L. Martí-Bonmatí and M. E. Brummer (2008). "k-Space tutorial: an MRI educational tool for a better understanding of k-space." Biomedical imaging and intervention journal 4(1).

Oguz, I., M. Farzinfar, J. Matsui, F. Budin, Z. Liu, G. Gerig, H. J. Johnson and M. A. Styner (2014). "DTIPrep: quality control of diffusion-weighted images." Frontiers in neuroinformatics 8: 4.

Ojala, T., M. Pietikainen and T. Maenpaa (2002). "Multiresolution gray-scale and rotation invariant texture classification with local binary patterns." IEEE Transactions on pattern analysis and machine intelligence 24(7): 971-987.

Pierpaoli, C., L. Walker, M. Irfanoglu, A. Barnett, P. Basser, L. Chang, C. Koay, S. Pajevic, G. Rohde and J. Sarlls (2010). "TORTOISE: an integrated software package for processing of diffusion MRI data." Book TORTOISE: an integrated software package for processing of diffusion MRI data (Editor ed^ eds) 18: 1597.

Rampasek, L. and A. Goldenberg (2018). "Learning from everyday images enables expert-like diagnosis of retinal diseases." Cell 172(5): 893-895.

Reuter, M., M. D. Tisdall, A. Qureshi, R. L. Buckner, A. J. van der Kouwe and B. Fischl (2015). "Head motion during MRI acquisition reduces gray matter volume and thickness estimates." Neuroimage 107: 107-115.

Revaud, J., G. Lavoué and A. Baskurt (2009). "Improving Zernike moments comparison for optimal similarity and rotation angle retrieval." IEEE transactions on pattern analysis and machine intelligence 31(4): 627-636.

Russakovsky, O., J. Deng, H. Su, J. Krause, S. Satheesh, S. Ma, Z. Huang, A. Karpathy, A. Khosla and M. Bernstein (2015). "Imagenet large scale visual recognition challenge." International journal of computer vision 115(3): 211-252.

Samala, R. K., H.-P. Chan, L. Hadjiiski, M. A. Helvie, C. Richter and K. Cha (2018). Cross-domain and multi-task transfer learning of deep convolutional neural network for breast cancer diagnosis in digital breast tomosynthesis. Medical Imaging 2018: Computer-Aided Diagnosis, International Society for Optics and Photonics.

Samala, R. K., H.-P. Chan, L. M. Hadjiiski, M. A. Helvie, C. Richter and K. Cha (2018). "Evolutionary pruning of transfer learned deep convolutional neural network for breast cancer diagnosis in digital breast tomosynthesis." Physics in Medicine & Biology 63(9): 095005.

Schenck, J. F. (1996). "The role of magnetic susceptibility in magnetic resonance imaging: MRI magnetic compatibility of the first and second kinds." Medical physics 23(6): 815-850.

Simmons, A., P. S. Tofts, G. J. Barker and S. R. Arridge (1994). "Sources of intensity nonuniformity in spin echo images at 1.5 T." Magnetic resonance in medicine 32(1): 121-128.

Simonyan, K. and A. Zisserman (2014). "Very deep convolutional networks for large-scale image recognition." arXiv preprint arXiv:1409.1556.

Smith, R., R. Lange and S. McCarthy (1991). "Chemical shift artifact: dependence on shape and orientation of the lipid-water interface." Radiology 181(1): 225-229.

Szegedy, C., W. Liu, Y. Jia, P. Sermanet, S. Reed, D. Anguelov, D. Erhan, V. Vanhoucke and A. Rabinovich (2015). Going deeper with convolutions. Proceedings of the IEEE conference on computer vision and pattern recognition.

Tajbakhsh, N., J. Y. Shin, S. R. Gurudu, R. T. Hurst, C. B. Kendall, M. B. Gotway and J. Liang (2016). "Convolutional neural networks for medical image analysis: Full training or fine tuning?" IEEE transactions on medical imaging 35(5): 1299-1312.

Van Dijk, K. R., M. R. Sabuncu and R. L. Buckner (2012). "The influence of head motion on intrinsic functional connectivity MRI." Neuroimage 59(1): 431-438.

Victoroff, J., W. Mack, S. Grafton, S. Schreiber and H. Chui (1994). "A method to improve interrater reliability of visual inspection of brain MRI scans in dementia." Neurology 44(12): 2267-2267.

Wang, J. and L. Perez (2017). "The effectiveness of data augmentation in image classification using deep learning." Convolutional Neural Networks Vis. Recognit.

Wood, M. L. and R. M. Henkelman (1985). "MR image artifacts from periodic motion." Medical physics 12(2): 143-151.

Xue, Y., R. Zhang, Y. Deng, K. Chen and T. Jiang (2017). "A preliminary examination of the diagnostic value of deep learning in hip osteoarthritis." PloS one 12(6): e0178992.